\documentstyle[aps]{revtex} 

\input epsf

\begin{document}


\preprint{WM-00-102; hep-ph/0002271}

\title{Scaling and Duality in Semi-exclusive Processes}

\author{ Andrei Afanasev{\footnotemark}}
\address{ Department of Physics, North Carolina Central University,
Durham,  NC 27707\\ and 
Thomas Jefferson National Accelerator Facility,
12000 Jefferson Avenue, Newport News, VA 23606}

\author{ Carl E. Carlson and Christian Wahlquist}
\address{ Nuclear and Particle Theory Group, Physics Department, College of
William and Mary, Williamsburg, VA 23187-8795 }


\footnotetext{\vglue -25pt $^*$On leave from Kharkov
Institute  of Physics and Technology, Kharkov, Ukraine.}


\date{printed \today
}

\maketitle

\begin{abstract}
{\small  We discuss extending scaling and duality studies to semi-exclusive
processes.  We show that semi-exclusive hard pion photoproduction should
exhibit scaling behavior in kinematic regions where the photon and pion both
interact directly with the same quark. We show that such kinematic regions
exist.  We also show that the constancy with changing momentum transfer of
the resonance peak/scaling curve ratio, familiar for many resonances in deep
inelastic scattering, is also expected in the semi-exclusive case.
 
\vglue -20pt}

\end{abstract}

\widetext \vglue -6.3cm  \hfill WM-00-102; hep-ph/0002271
          \vglue 5.8cm 




 \section{Introduction}


Scaling is a well established phenomenon in deep inelastic scattering (DIS). 
The cross section with specific kinematic factors removed gives structure
functions that depend on only the scaling variable $x_B$, up to calculable
logarithmic corrections.  In addition, an inclusive-exclusive
connection---``Bloom-Gilman scaling''~\cite{bg}---is observed in these totally
inclusive (at least on the hadronic side) reactions.  Duality in this
situation means that resonance bumps observed in the structure functions at
low momentum transfers $Q^2$ average out to the smooth structure function
measured at higher momentum transfers but the same $x_B$.  Usually, but not
always, duality is realized in such a way that as the resonance peak moves in
$x_B$ with changing $Q^2$, the ratio of the peak height to the height of the
scaling curve evolved from higher $Q^2$ is constant.

Both scaling and scaling violation have played a crucial role in
understanding the constituents of elementary particles and in establishing
QCD as as the accepted theory of the strong interactions.  Duality is in detail
less well understood~\cite{dgp,cm90}.  It seems, however, to show that the
fundamental single quark QCD process is still decisive in setting the scale
of the reaction in the resonance region, and that the crucial role of the
final state interactions in forming the resonance becomes moot when averaged
over, say, the resonance width.  This last observation, if reliably
understood,  could allow one to use duality to study the structure functions
in the interesting and still experimentally uncertain $x_B \rightarrow 1$
region.  For a fixed available energy, $x_B \rightarrow 1$ means getting into
the resonance region and if one were sure of the connection of the resonance
region average to the scaling curve, one could determine the scaling curve
significantly closer to the kinematic upper endpoint.

Departing from DIS, we want to continue test our ability to understand and
apply QCD to describe hadronic processes.  A set of processes that can
be a new testing ground for both scaling and duality phenomena are
semi-exclusive reactions typified by 
\begin{equation}
\gamma + p \rightarrow \pi + X,
\end{equation}
where the photon may be real or virtual.  These processes are the topic of
this paper.  We shall study suitable kinematic variables for the general case
and, when we get more detailed, give special attention to photoproduction with
large photon to pion momentum transfer, $t$.

A first requirement is to find a scaling region.  This problem has been
studied in the high $Q^2$--low $(t/Q^2)$ limit, focusing on the totally
exclusive reaction but with extension to the semi-exclusive
case~\cite{dombeywest}.  These authors found that scaling functions would
exist, provided the photon and pion currents directly and successively
interacted with the same quark while the rest acted as spectators.  

\begin{figure} 

\centerline {\epsfysize 1.7 in \epsfbox{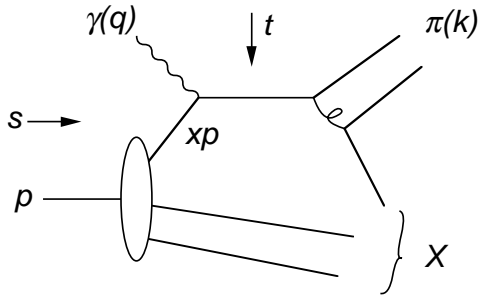}  }

\caption{Direct (also isolated or short-distance) pion production, with some
kinematics indicated.} 
\label{direct}

\end{figure}

We here, concentrating on photoproduction at high $|t|$, show that
perturbative QCD (pQCD) predicts there is indeed a scaling region.  We shall
below show the kinematic factors that connect the cross section to the
expected scaling function.  We shall also see that the scaling region does
require kinematics where photopion production is dominated by direct
interactions of both the photon and the pion~\cite{cw93,acw98,bdhp99}, such
as seen in Fig.~\ref{direct}.   In particular, one must avoid regions where
the pion comes from soft processes or comes as part of a jet from a
fragmenting parton.  In earlier work we were able to show that regions of
direct pion production exist, and therefore there are regions where we can
find a scaling function.

When scaling is established at high $|t|$, one can study duality. One can ask
whether the scaling curve from high $Q^2$ or $t$ a decent average over the
resonance bumps seen at the same $x$ but lower $Q^2$ or $t$?  Duality in this
sense appears to be true for all the resonances seen in DIS.  Further, one
can ask if the bump to continuum ratio is constant as $Q^2$ or $t$ changes? 
This constancy is seen in DIS for most resonances, but not for the
$\Delta$(1232).  While studying the kinematics and working in photoproduction context, we will see that it is possible in a single
experiment with good kinematic coverage to probe a given $x$ region over a
wide range of $m_X$ from the resonance region to well into the continuum
region.

The paper proceeds as follows.  Section~\ref{kinematics} will discuss the
kinematics and scaling variable for  the semiexclusive process. 
Section~\ref{regions} will show how a scaling function emerges for
semi-exclusive hard pion photoproduction, and also show the existence of a
region where direct pion production dominates, specifically for a situation of
30 GeV incoming photons.  Section~\ref{bumps} will show that pQCD expectations
for the resonance peak/scaling curve ratio at changing $|t|$ are similar to
what one sees in DIS.  Section~\ref{end} will offer some conclusions.


\section{kinematic variables}                   \label{kinematics}


For the process $\gamma + p \rightarrow \pi + X$, define the Mandelstam
variables by
\begin{equation}
s = (p+q)^2, \quad t = (q-k)^2, \quad u = (p-k)^2 \ .
\end{equation}
Define $x$ in general by,
\begin{equation}
x = {-t \over s+u- 2m_N^2 - q^2 - m_\pi^2} ,
\end{equation}
and note that all quantities defining $x$ are experimentally
measurable~\cite{cw93,acw98,bdhp99}.  One can show $0 \leq x\ \leq 1$, and
$x = 1$ corresponds to the case that $X$ is a nucleon.  Also generally, the
hadronic mass recoiling against the pion is given by
\begin{equation}
m_X^2 = m_N^2 -t \left({1\over x} - 1 \right).
\end{equation}

Specializing to the case where direct pion production, Fig.~\ref{direct}, is
the underlying process, in the limit of high $t$ or $s$ and high
recoil mass $m_X$,  one can show that this $x$ is the fraction of the target's
momentum carried by the struck quark.  The proof involves defining
Mandelstam variables for the subprocess $\gamma + q \rightarrow \pi + q'$.
We anticipate the result by letting the momentum of the struck quark
be called $xp$, and get
\begin{eqnarray}   \label{submandelstam}
\hat s &=& (xp + q)^2 = x (s - q^2) + q^2 \ , \nonumber \\
\hat t &=& (q  - k)^2 = t  \ ,\nonumber \\
\hat u &=& (xp - k)^2 = x u \ ,
\end{eqnarray}
where we have neglected masses but not $q^2$.  For the direct pion production
subprocess,
\begin{equation}
\hat s + \hat t +\hat u = q^2 \ ,
\end{equation}
and substituting Eqn.~(\ref{submandelstam}) leads to the identification of $x$
the momentum fraction to
$x$ the experimental observable, when individual particle masses can be
neglected. 

Thus $x$ is a precise analog of the observable
$x_B = Q^2/2m_N \nu$ in deep inelastic scattering. 
Our formulas should and do connect to the well known ones for
deep inelastic kinematics in the limit $k \rightarrow 0$.  In this limit, 
$u \rightarrow m_N^2$ and $t \rightarrow q^2 \equiv -Q^2$ and
\begin{eqnarray}
m_X^2 &=& m_N^2 + Q^2 \left({1\over x} - 1 \right) \ , \nonumber \\
x &=& {-t \over s -m_N^2 -q^2} = {Q^2 \over 2 p \cdot q} \ ,
\end{eqnarray}
without approximation.

Still regarding deep inelastic scattering, Bloom and Gilman~\cite{bg} found
that near threshold scaling worked better if one used a revised variable
defined as 
$1/x'_B = 1/x_B + m_N^2/Q^2$. By analogy to Bloom and Gilman's proposal we
could define a modified scaling variable with $-t$ replacing $Q^2$: 
\begin{equation}                                   \label{modified}
{1\over x'} = {1\over x} + {m_N^2\over -t},
\end{equation}
whence
\begin{equation}
m_X^2 = -t {1-x' \over x'}.
\end{equation}
One should keep this possibility in mind here also.

Another situation, related to the one we are pursuing, is 
semi-exclusive deep inelastic scattering with parallel kinematics.  This means
high $Q^2$ and an observed meson with three-momentum parallel to the incoming
photon, in the lab.  In this case, there is a variable $z$ defined by
\begin{equation}
z \equiv {p \cdot k \over p \cdot q} \ .
\end{equation}
and obtain
\begin{eqnarray}
m_X^2 &=& m_N^2 + Q^2  (1-z)  \left({1\over x} - 1 \right) \ , \nonumber \\
x &=&  {Q^2 \over 2 p \cdot q} \ ,
\end{eqnarray}
with the neglect of terms of ${\cal O}(m_N^2 x^2 / Q^2)$.


\section{scaling and kinematic regions}              \label{regions}


Now we shall focus on hard pion photoproduction, where $q^2 = 0$, $k_T$ is
large, and $|t|$ is large.

We will be mainly interested in direct pion production with $m_X$ large,
and in the transition to the exclusive reactions 
$\gamma+N \rightarrow \pi X$.  Other processes do contribute.  In particular
there are soft processes, and processes where the pion is produced as part of
the fragmentation of a quark or gluon into a
jet.  These processes can be evaded if one can go to sufficient transverse
momentum.  We will comment on them briefly before proceeding.

\begin{figure} 

\centerline {\epsfysize 1.7 in \epsfbox{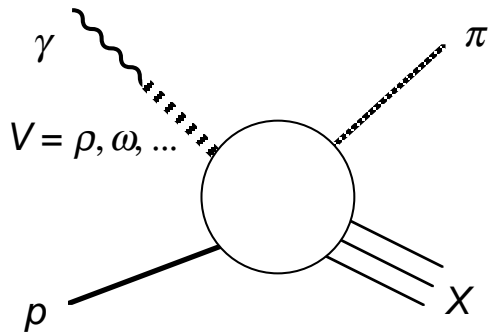}  }
\vglue 2mm

\caption{Soft processes, approximated by vector meson dominance.} 
\label{soft}

\end{figure}

Soft processes are frequently approximated using vector meson dominance of
the photon interaction, illustrated in Fig.~\ref{soft}.  They are important
at low transverse momenta, although the boundary between ``low'' and ``high''
is higher than one might expect, namely around 2 GeV.  We have considered
these processes in a fashion suitable for the present context
in~\cite{acw00}; one can also find a representation of them in
PYTHIA~\cite{torbjorn}.


\begin{figure} 

\centerline {\epsfysize 1.7 in \epsfbox{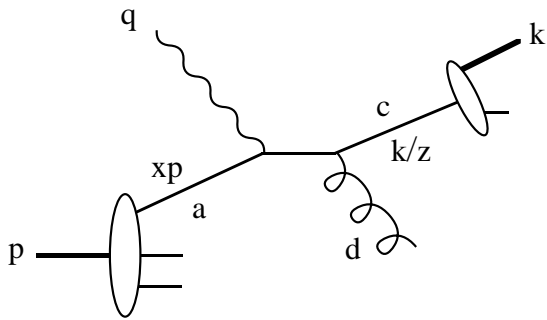}  }
\vglue 2mm

\caption{A fragmentation process, where the pion is labeled as having
momentum $k$.} 
\label{fragmentation}

\end{figure}

Moderate transverse momenta hard pions can be produced by a fragmentation of
a parton. The process is perturbatively calculable and could be a way to learn
about polarized of unpolarized gluon distributions of the
target~\cite{acw98,many}; one example is illustrated in
Fig.~\ref{fragmentation}.  


Neither the fragmentation nor the soft process is useful for the present
duality study.  The reason is that the experimental $x$ variable for them does
not have a unique connection to the quark momentum fraction, and we will not
be able to prove a scaling relation for them.  

Direct pion production, however, does have the nice connection between an
experimentally observable $x$ and the struck quark momentum fraction, and it
is calculable in pQCD.  It is a higher twist process.  Factors of the decay
constant enter the amplitude, representing the quark-antiquark wave function
of the pion at the origin, and must be dimensionally compensated by an extra
power of $s$ in the cross section.   Nonetheless, it can dominate over
fragmentation at high
$k_T$ because it always gives all the transverse momentum in the pion
direction to the one pion.  For the direct process, we
can operationally define a scaling function $F(x,t)$ by
\begin{eqnarray}                                      \label{scaling}  
E_\pi {d\sigma \over d^3k} &=& 
    {(s - m_N^2) x^2 \over - \pi t}  {d\sigma \over dx \, dt}
\nonumber \\
&=&
    {(s - m_N^2) x^2 \over - \pi t}  \ 
    {d\hat\sigma \over dt} (\gamma q \rightarrow \pi q') \ 
    F(x,t)   \ .
\end{eqnarray}
I.e., the scaling function is related to the cross section by some kinematic
factors, which are partly explicit above and partly given in terms of the
cross section for the subprocess
\begin{eqnarray}                                      \label{subproc}
{d\hat\sigma(\gamma q \rightarrow M q') \over dt} &=&
  {128 g_F^2 \pi^2 \alpha \alpha_s^2 \over 27 (-t) \hat s^2}
    I_M^2
  \left({e_q \over \hat s} + {e_{q'} \over \hat u} \right)^2
     \nonumber \\[1ex] &\times&
\left[ \hat s^2 + \hat u^2
                 + \lambda h
       \left( \hat s^2 - \hat u^2
       \right)
    \right],
\end{eqnarray}
where we should substitute quark charges relevant for pion being produced,
for example $e_q = e_u$ and $e_q' = e_d$ for the $\pi^+$.    The flavor
factor $g_F$ is unity for the
$\pi^\pm$ and $1/\sqrt{2}$ for the $\pi^0$.  $I_M$ is for the present purpose
a constant factor, but if the perturbative calculation is valid, it will be
given in terms of the distribution amplitude of the meson as 
$\int d\xi_1 \, \phi_M(\xi,\mu^2) / \xi_1$.  For the asymptotic distribution
amplitude, $I_\pi = \sqrt{3} f_\pi/2$ with $f_\pi \approx 93$ MeV.

We have included polarization dependence for future use:
$\lambda$ is the helicity of the photon and $h$ is twice the helicity of the
target quark. Of course, duality can be tested with polarization as well as
without.

The reason to believe that the above expression, Eqn.~(\ref{scaling}),
produces a scaling formula is that  the perturbative formula valid for short
distance pion production (the process of Fig.~\ref{direct}) is, 
\begin{eqnarray}
E_\pi {d\sigma \over d^3k} &=& 
    {(s - m_N^2) x^2 \over - \pi t} \nonumber \\ 
     &\times&  \sum_q
    {d\hat\sigma \over dt} (\gamma q \rightarrow \pi q')  \,
     G_{q/T}(x,\mu^2) 
\end{eqnarray}

Thus where perturbation theory works, there is a
scaling function $F(x,t)$ is mainly dependent on $x$. We can
relate it to the quark distributions (with weak dependence on
the scale $\mu^2$, which we may set to $t$), as in DIS.  We expect the
formulas will be mainly applied in the high $x$ region, where valence quarks
dominate.  Hence the comment on the choice of the quark charges just above.

Let us comment on the fact that the presence of a hard gluon exchange (see
Fig.~\ref{direct}) indicates that one needs sufficiently high energies to
apply the pQCD formalism.  However, since only one pion distribution
amplitude is involved for the direct process, if the photon attaches to the
produced $q \bar q$ pair of Fig.~\ref{direct} (the worse case), the average
virtuality of the gluon in question corresponds to the one determining the
pion electromagnetic form factor at $Q^2 \approx 20 (35)$ GeV$^2$ scale, for
the asymptotic (Chernyak-Zhitnitsky) pion distribution amplitude assuming a
CEBAF energy of 12 GeV, pion emission angle of 22$^\circ$, and $m_X = 2$ GeV
(see Ref.~\cite{acw98} for details).  Therefore one may hope to observe a
single-gluon exchange, which is a higher twist effect, in inclusive
photoproduction of pions even at CEBAF energies generally considered not
high enough to reach the perturbative QCD domain.  Indications of direct pion
production off a quark were obtained in $\pi N$ scattering (see
Ref.~\cite{owens} for references and discussion).

We may now ask if this scaling function dual, in a Bloom-Gilman like
sense, to the bumpier curve one will get in the resonance
region?  The resonance region, of course, is what we have at the very highest
transverse momentum, where there is very little energy left over to put into
recoil mass.

Formally, the duality relation then may be written as an integral of
the differential cross section $d\sigma/dx\, dt(\gamma N \rightarrow \pi X)$
and in the region of the direct process dominance reads 
\begin{eqnarray}                                               \label{dual}
\int_{(1-{m_X^2 - m_N^2 \over t})^{-1}}^1 &dx& \,
     \sum_q G_{q/N}(x) {d\sigma \over dt}(\gamma q \rightarrow \pi q')
                                  \nonumber \\
\propto &\sum_R&
     {d\sigma \over dt}(\gamma + N \rightarrow \pi + R)  \ .
\end{eqnarray}
Summation in the right hand side of Eqn.~(\ref{dual}) is done over all
resonances
$R$ with masses $m_R \leq m_X$, with the nucleon final state included.  If the
parton distribution function of the nucleon is $G_{q/N} \sim (1-x)^3$
at $x \rightarrow 1$ and the subprocess $\gamma q \rightarrow \pi q'$
cross section is determined by the one-gluon exchange mechanism of
Fig.~\ref{direct}, then as will be shown in the next section duality as in
Eqn.~(\ref{dual}) requires that the resonance excitation cross section
${d\sigma / dt}(\gamma + N \rightarrow \pi + R) \propto 1/s^7$ at fixed
$t/s$---the result known from the constituent counting rules~\cite{count}. 
The duality relation above could also be written using the modified
scaling variable $x^\prime$ from Eqn.~(\ref{modified}).

One should ask if the proper regions exist. There needs to be a region where
direct pion production dominates, where one can measure the scaling curve and
see how it tails off into the resonance region.  Such a region does exist. 
>From earlier studies~\cite{acw98,acw00} we have the machinery to calculate the
direct pion and fragmentation process, and estimate the VMD processes, and
have shown that the direct process, even though it is higher twist, does take
over at some point if we have enough initial photon energy.  

\begin{figure} [ht]  

\hglue -.3in {\epsfxsize 3.5 in \epsfbox{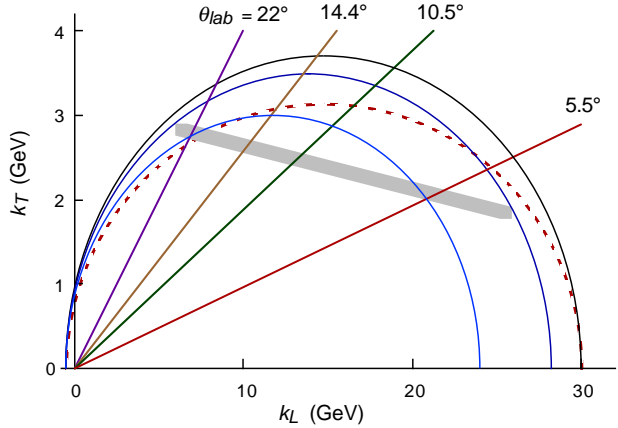}  }

\vglue -1.2in

\caption{Kinematic regions relevant to studying duality in $\gamma N
\rightarrow \pi X$ for 30 GeV incoming photons.  The upper solid elliptical
line is the overall kinematic limit, and corresponds to
$\gamma N \rightarrow \pi N$.  The other two solid elliptical lines have
$m_X$ of 2 and 3.5 GeV, respectively. Between the two upper solid elliptical
lines lies the resonance region. Above the grey line soft processes and
fragmentation processes are small, and direct pion production dominates. 
Hence the region between the grey line and the middle elliptical line is the
region that can be smoothly compared to the resonance region. The dotted
elliptical curve is for $x$ fixed at 0.7.} 
\label{fig1}

\end{figure}

A useful presentation of our calculated results is shown in
Fig.~\ref{fig1}.  The figure attempts to show that we can follow a given $x$
region from the resonance region until well into the scaling region, and do
so in a single experiment.  The axes of Fig.~\ref{fig1} give the outgoing
pion transverse and longitudinal momenta, in the target rest frame.  Some
labeled straight lines give the pion angle relative to the incoming beam.  The
three solid elliptical curves each correspond to a fixed value of recoil mass
$m_X$.  The outermost curve has $m_X = m_N$ and thus corresponds to the
quasielastic process 
$\gamma N \rightarrow \pi N$, and also marks the kinematic limit of pion
momenta.  The next curve has $m_X = 2$ GeV, and the innermost solid curve has
$m_X = 3.5$ GeV.  Thus the region between the two outermost curves is the
resonance region, and the region within the middle solid curve is the
continuum region.  The segment above the grey band is the region where direct
pion production dominates.  For us, this is the ``good region.''  (As a side
note, the grey band is straighter than we might have guessed, especially since
it is made up of two parts.  The central part comes from the fragmentation
process growing larger.  Both ends come from VDM, as modeled in an earlier
note, which we think was conservative in estimating the size of the VDM
contributions.)  Finally comes the important dotted elliptical curve,
which has a constant $x$, specifically $ x= 0.7$ in this case.  We see that
we can thinkably measure the putative scaling function in the resonance region
at small pion angles, and then by moving to larger angle, follow its behavior
at the same $x$ but larger $m_X$ (and larger $|t|$) well out of the
resonance region, before running into a region where fragmentation or soft
processes dominate. 


(As another aside, lines of constant $|t| \ne 0$ on this plot would be
parabolas opening to the right, and passing through the small line segment
between the origin and the lower (negative) limit of $k_L$; $|t| =
0$ occurs along the positive $k_L$ axis.)


\section{resonance bumps vs.\ the scaling curve}    \label{bumps}


There is always a resonance region.  In plots of $F(x,t)$ vs. $x$, the bumpy
resonance region slides to the right with increasing $|t|$. In the
corresponding DIS case the bumps slide neatly down the curve, with the
resonance/smooth curve ratio observed to stay the same, for most resonances. 
Within pQCD, this is expected theoretically~\cite{cm90} as a consequence of
the known behaviors of the scaling curve as $x \rightarrow 1$ and the
predicted falloff of the resonance transition form factors at high $Q^2$. We
can show that the resonance/continuum constancy is consistent with pQCD in
the semi-exclusive case also.

We need to find the behavior of
\begin{eqnarray}
F_{res}(x,t) = {d \sigma \over dx\, dt} (\gamma N \rightarrow \pi R)
/
{d \sigma \over dt} (\gamma q \rightarrow \pi q')
\end{eqnarray}
at (say) the resonance peak for large $|t|$ (and $x \rightarrow 1$).  The
denominator in this limit is
\begin{equation}
{d \sigma \over dt} = g(t/s) |t|^{-3}
   \propto (1-x)^3
\end{equation}
where $g(t/s)$ is a known function (see Eqn.~(\ref{subproc})) which does not
go to zero for $t/s$ finite, and we have used $1/t \propto (1-x)$ in the
stated limit.

The numerator for a finite width resonance can be approximated by (for 
$x \rightarrow 1$),
\begin{eqnarray}
\left(  d\sigma \over dx \, dt \right)_{res} &\approx&
{|t| \over 2 m_X} \left( d\sigma \over dm_X \,dt \right)_{res}
                                   \nonumber \\
&\approx& {|t| \over 2 m_R} \left( d\sigma \over dt \right)_{res}
   {\Gamma / 2\pi \over (m_X-m_R)^2 + \Gamma^2/4}
\end{eqnarray}
where $\Gamma$ is the width of the resonance and we have used a simple
lorentzian form to give the resonance shape.  The pQCD scaling
rules~\cite{count} tell us that
\begin{equation}
\left( d\sigma \over dt \right)_{res} = f(t/s) |t|^{-7}
\end{equation}
where $f(t/s)$ is not known but in general it should not go to zero for
finite $t/s$.  Thus,
\begin{eqnarray}
\left( d\sigma \over dx \, dt \right)_{res\ peak}
\approx {1\over \pi m_R \Gamma}  f(t/s) |t|^{-6} \propto (1-x)^6
\end{eqnarray}
Thus,
\begin{equation}
F_{res\ peak} (x,t) \propto (1-x)^3  \ .
\end{equation}
This is how the height of a resonance peak fall with $x$ as 
$x \rightarrow 1$.  It is also precisely the pQCD expectation for the scaling
curve. Hence the resonance/continuum ratio is in general constant, at least at
high $|t|$, as it is for DIS.

In DIS, the Delta(1232) is an
exception, as it falls markedly with $Q^2$~\cite{s,cm93}; $Q^2$ in lepton
scattering is the analog of $-t$  in hard meson photoproduction.  It will be
interesting to see if the Delta(1232) disappears with increasing $|t|$ and if
the, say, S$_{11}$(1535) stays up at high $|t|$.  Recall that in pQCD, the
disappearing Delta in electron scattering is explained as an accident having
to do with the specifics of the Delta and nucleon wave functions~\cite{cp88}. 
We should not expect this to be necessarily replicated in pion photoproduction
since the integrals over the distribution amplitudes will involve different
weightings.


\section{Conclusions and Discussion}                    \label{end}


Semi-exclusive processes give an opportunity to extend the studies of scaling
and duality, which in deep inelastic scattering have been
fruitful in verifying our understanding of QCD and in pushing our effort to
deepen that understanding.  

It appears that scaling in the sense that the cross section is directly
related to a scaling function that depends, up to logarithmic corrections, on
just one variable.  The scaling variable for semi-exclusive processes, given
in the text, is related to the momentum fraction of the struck quark, just
like the scaling variable in deep inelastic scattering.  However,
scaling, at least as we have been able to present it in this paper, works in
semi-exclusive process only when the pion is produced directly off the same
quark that absorbs the incoming photon.  We have been able to show,
theoretically, that such a scaling region does exist.  

One should bear in mind that there are soft kinematic regions where one does
not know where the pion comes from, and fragmentation regions where the pion
is produced at some remove from the fundamental process that initiates the
reaction.  We do not know of a scaling function for these regions, and it is
not trivial that one can avoid them, but one can.  One should also bear in
mind that a certain amount of initial energy is needed to be able to produce
a scaling region.  For incoming photon lab energy 16 GeV or below and our
present estimates of the vector meson dominance contributions, it does not
appear that there is a region where VMD is not the biggest process for
photoproduction, at least if one does not make any additional cuts.  However,
a there are possibilities for reducing the necessary incoming energy.  One
follows from noting that a directly produced pion is also a pion produced
in kinematic isolation, not as part of a jet, and one can consider an
``isolation cut,'' a requirement that there be no other particles collinear
with the pion.  Another possibility is to have the photon off shell, since
then the vector meson propagator is significantly reduced, reducing the VMD
contributions without there being an equal reduction for other contributions.
We are hopeful that using electroproduction and isolation cuts can make the
incoming energy requirement low enough to fit an upgraded CEBAF range, but
are deferring detailed elaboration.

The existence of a scaling region also allows one to consider the
inclusive-exclusive connection with the resonance region.  Will the resonance
bumps average out to the smooth scaling curve measured at higher $-t$ and
evolved to lower $-t$~\cite{nonevolution}?   Will the resonance peak to
scaling curve ratio be independent of $t$?  In deep inelastic physics, it
does appear that the final state interactions which produce the resonance are
irrelevant to the overall rate of resonance region production, if one does a
suitable average.  And we have shown that for the semi-exclusive case, as in
the deep inelastic case, the resonance to continuum ratio should be constant,
barring special circumstances.  A special circumstance in the deep inelastic
case occurs for the $\Delta(1232)$, which disappears into the scaling curve
with increasing $Q^2$.   One would like to know if similar phenomena occur in
other situations.

Testing scaling and duality in inclusive photoproduction of mesons requires
coverage of the large-$x$ region, where the cross sections are rapidly falling
as $x$ approaches it upper limit.  Therefore such a uniquely designed
high-luminosity machine as CEBAF, with a bit more energy, could do an
excellent job in these duality studies.

We thank Nathan Isgur, Wally Melnitchouk, Chris Armstrong, Rolf Ent, and
Cynthia Keppel for useful discussions about duality and give the latter three
a second thanks for showing us their experimental duality results from CEBAF.
AA thanks the US Department of Energy for support under contract
DE-AC05-84ER40150; CEC and CW thank the NSF for support under grant
PHY-9900657.

%
%
%
%
%

\vglue -12pt


\end{document}